\documentclass[review,number,sort&compress]{elsarticle}

\usepackage{amssymb}
\usepackage{amsmath}
\usepackage{gensymb}

\usepackage{wasysym}

\usepackage{graphics} 
\usepackage{graphicx}
\usepackage{color}
\usepackage{subfigure}

\usepackage{rotating}
\usepackage{textcomp}
\usepackage{subfigure}
\usepackage{xcolor}

\usepackage{amsmath,amsthm,amssymb,amsfonts}
\usepackage{units}
\usepackage{hyperref}

\usepackage{changepage}

\usepackage{siunitx}
\usepackage{isotope}
\usepackage{tikz}
\usepackage{afterpage}
 
\usepackage{dcolumn}

\usepackage{lineno}

\usepackage{verbatim}
\usepackage{here}

\journal{Nuclear Instruments and Methods in Physics Research A}

\begin{document}
\begin{frontmatter}
\title{
An ultracold neutron storage bottle for UCN density measurements}

\author[PSI]{G.~Bison}
\author[PSI]{F.~Burri}
\author[PSI]{M.~Daum}
\author[PSI,ETH]{K.~Kirch}
\author[ETH]{J.~Krempel}
\author[PSI]{B.~Lauss\corref{cor1}}
\author[PSI]{M.~Meier}
\author[PSI,ETH]{D.~Ries\corref{cor2}}
\author[PSI]{P.~Schmidt-Wellenburg}
\author[PSI]{G.~Zsigmond}


%
\address[PSI]
{Paul Scherrer Institute (PSI), CH-5232 Villigen PSI, Switzerland}
\address[ETH]
{Institute for Particle Physics, Eidgen\"ossische Technische Hochschule (ETH), Z\"urich,  Switzerland}

\cortext[cor1]{Corresponding author, bernhard.lauss@psi.ch}
\cortext[cor2]{Corresponding author, dieter.ries@psi.ch}



\begin{abstract}
We have developed a storage bottle
for ultracold neutrons (UCN) in order to 
measure the UCN density at the beamports of 
the Paul Scherrer Institute's (PSI) UCN source.
This paper describes the 
design, construction and commissioning of the
robust and mobile storage bottle with a 
volume comparable to typical storage experiments (\SI{32}{\liter})
e.g. searching for an electric dipole moment of the neutron.
\end{abstract}



\end{frontmatter}





\section{Introduction}

Ultracold neutrons (UCN) are 
neutrons with kinetic energies below about 350\,neV, 
corresponding to a temperature of about 4\,mK. 
Such neutrons are reflected under all angles of incidence 
from surfaces of suitable materials.
Therefore, UCN can be confined in material containers,
dubbed 'storage bottles', in which
storage times of several hundreds of seconds
are possible and 
ultimately limited by the neutron beta-decay lifetime of 880\,s.
This unique feature makes UCN ideal to study 
fundamental properties of the neutron 
in precision experiments \cite{Golub1991}
like
most prominently in the search for an electric dipole moment of the neutron 
(nEDM)
\cite{ramsey1982,Baker2006,Baker2011,SerebrovEDM2015,Pendlebury2015}
or 
the determination of the lifetime of the free neutron
\cite{Serebrov2005tau,Wie2011},
the measurement of neutron decay correlations
\cite{Liu2010,Plaster2012,UCNA2013}
or the investigation of neutron states bound in the Earth's gravitational field
\cite{GRANIT2002,Jenke2011,Jenke2014}.
The precision of such experiments is limited by the number of
UCN available to the experiment. Hence, worldwide efforts to provide
higher intensities of UCN are under way.
In addition to the ILL PF2 facility \cite{Steyerl1986}, 
which has been leading the field during the last 30 years,
several new UCN sources are already 
in operation 
\cite{LANL2013,Piegsa2014,Frei2007,Lauss2014}.
%

At the 
Paul Scherrer Institute (PSI), Villigen, Switzerland
a new UCN source was constructed and has been in operation 
since 2011
\cite{Lauss2014,Anghel2009,Lauss2011,Lauss2012}.

One important benchmark parameter for the
performance of a UCN source is the available 
UCN density which can be stored in an external experiment.
In order to determine the UCN densities at
different beamports or sources
in a consistent way, we have constructed a mobile storage vessel.

\section{Design and construction of the storage experiment}


In a first attempt,
using the prestorage vessel from Ref.~\cite{Blau2016},
which is a NiMo-coated glass tube,
and two beamline shutters,
UCN densities larger than 20\,UCN/cm$^3$ were measured
at PSI's West-1 beamport~\cite{Goeltl2012}.
However, these measurements were 
limited by UCN shutter actuation times and 
UCN losses during actuation.

The main goals for a new storage setup were:
construction using commercially available stainless steel (SST) tubing, 
mobility of the setup
and hence a necessary certain robustness of construction,
fast shutters without unnecessary UCN losses via openings
during movement.
The overall volume of the setup
should be similar
to a typical storage vessel
employed in a nEDM measurement,
i.e.~larger than \SI{20}{\liter},
and have 
a Fermi potential ($V_F$) on the surface 
well suited for UCN storage.
The employed stainless steel alloys have a 
calculated $V_F$ of about 188\,neV.
The Fermi potential in the 
UCN storage chamber of the currently best nEDM experiment 
is given by 
``deuterated polystyrene'' used as 
coating of the electrical insulator
($V_F$=161\,neV~\cite{Bodek2008}).

\subsection{UCN shutters}

\begin{figure}[h]
\begin{center}
\includegraphics[width=0.49\textwidth]{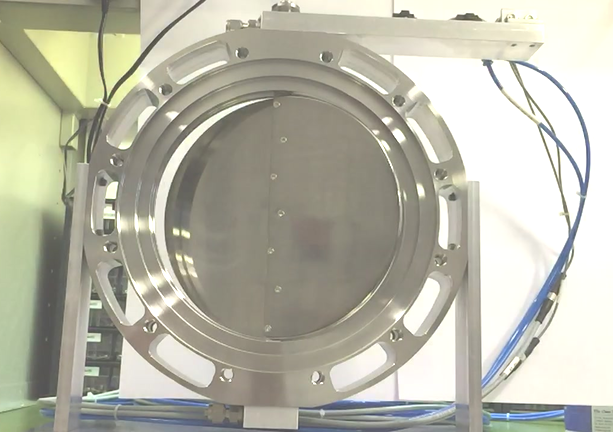}
\caption{Photograph of one of the shutters. Pressurized air 
and cables for position indicators
are on the right hand side. 
Cut-outs from the stainless steel body are for weight reduction.
The opening for UCN has a diameter of 200\,mm.
}
\label{shutterPhoto}
\end{center}
\end{figure}

Two butterfly-valve-type shutters
with \SI{200}{\milli\meter} opening were 
designed and built. 
One is depicted in Fig.~\ref{shutterPhoto}.
All parts inside the storage volume 
which can be in contact with UCN are made from stainless steel. 
The shutters can be connected to outer diameter 204\,mm standard tubes.
The shutter blades (SST type 1.4301) 
were polished mechanically and are mounted to the
rotating axis, 
such that all screws are mounted from one side, 
while the other side has a continuous surface.
The shutter housing material is SST type 1.4435.

\begin{figure}[h]
 \begin{center}
 \includegraphics[width=0.49\textwidth]{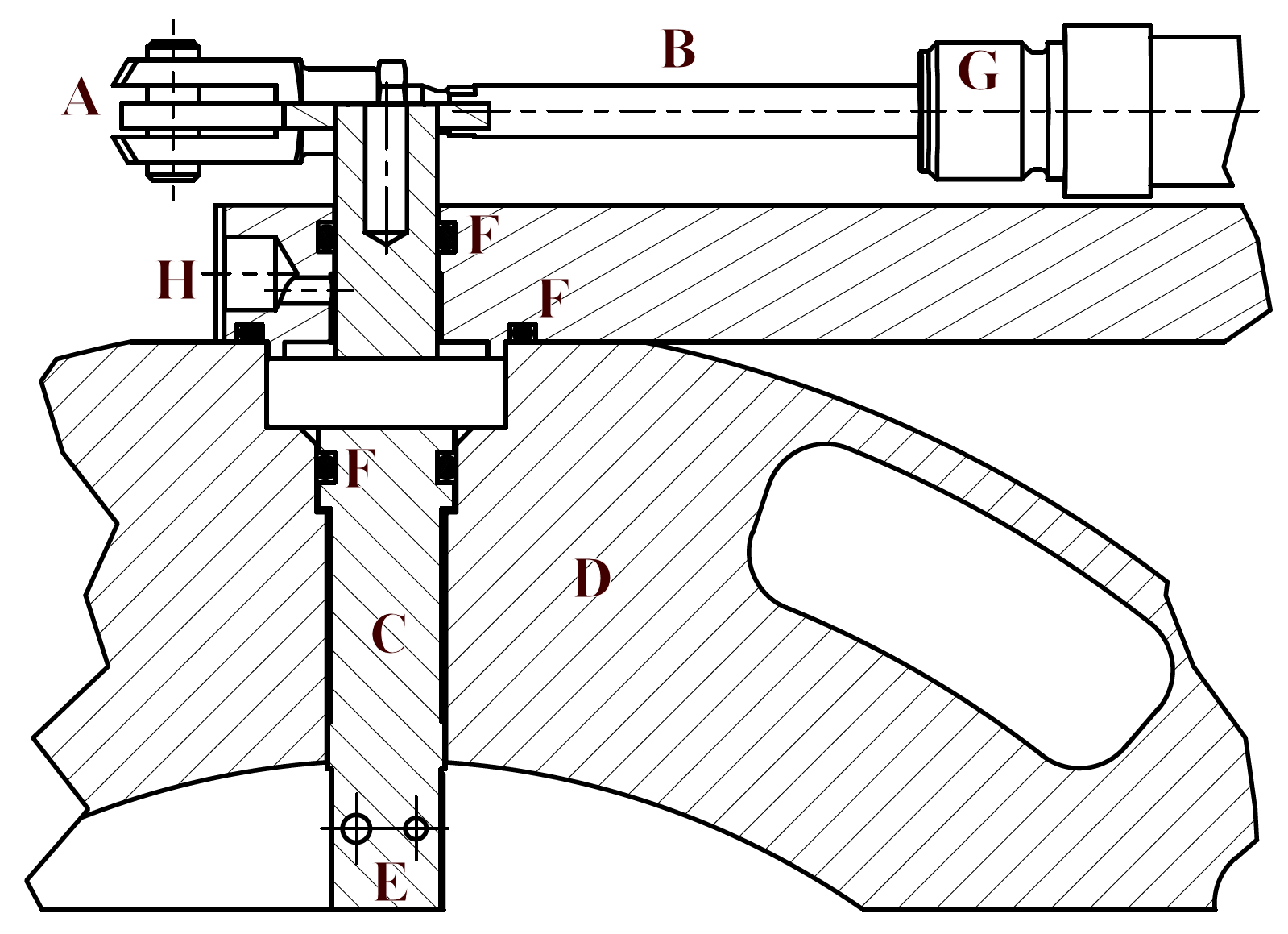}
\caption{UCN shutter: Technical drawing of the mechanical feedthrough. 
The pneumatic cylinder is acting on the shutter axis via a lever arm outside
the vacuum. The axis feedthrough can be differentially pumped. The holes
on the bottom of the axis are used to mount the shutter blades.
A: lever arm,
B: piston,
C: rotation axis,
D: frame,
E: axis mounting for blades,
F: viton sealing,
G: pneumatic actuator,
H: pump port;
}
\label{shutterDiffPumpCAD}
\end{center}
\end{figure}

The shutter disc can be rotated with  
a pneumatic linear actuator (Festo DSNU-16-50-PPV-A), 
which delivers a force greater than \SI{100}{\newton} along
a stroke of \SI{50}{\milli\meter}.
An air pressure of about \SI{6}{\bar} is used.
The mechanical feedthroughs can be differentially pumped through additional
tubes terminated by a KF16 flange.
Figure~\ref{shutterDiffPumpCAD} shows a technical drawing of the
differentially pumped mechanical feedthrough and actuation mechanism.

On average,
the shutter housings add each an average length of \SI{10.0\pm0.5}{\milli\meter} 
to the storage volume of the tube when the shutters are closed.

\begin{figure}[ht]
 \centerline{
  \includegraphics[width=0.79\textwidth]{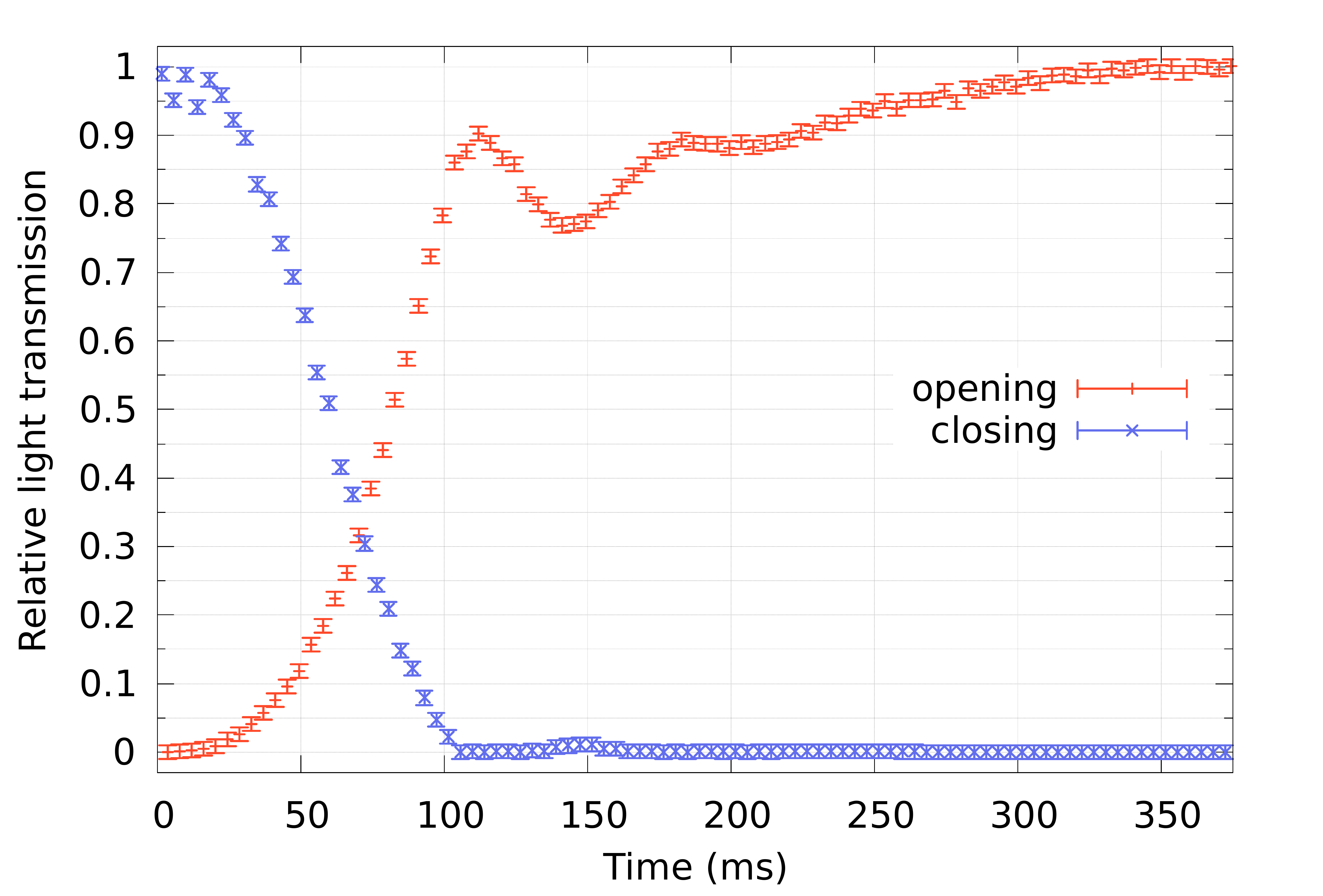}}
\caption{Opening and closing behavior of the UCN shutters measured by comparing
the visible white area of the background behind the shutter with the area visible
in fully opened position. The movements start at time=0.
Closing from fully open
to \SI{95}{\percent} closed takes less than \SI{100}{\milli\second}.
Opening from zero to \SI{90}{\percent} open takes
about \SI{120}{\milli\second}.
There is a small swing-back motion
caused by the break of the pneumatic cylinder. 
}
\label{OpeningClosing}
\end{figure}

Opening and closing times of less than
\SI{100}{\milli\second} were 
measured using a series of high speed camera pictures
as shown in Fig.~\ref{OpeningClosing}.

Using a duration $t$ of the movement of \SI{100}{\milli\second} and a radius
$r$ of the bottle of \SI{100}{\milli\meter}, the maximum
speed of the outer edge of the shutter blade during movement is 
\SI{1.57}{\meter\per\second}, which is considerably slower than the 
UCN velocities in the bottle and hence has 
only a small influence on the stored UCN density.

\subsection{Storage bottle}

A \SI{1000.0\pm0.5}{\milli\meter} long stainless steel tube, 
ID=\SI{200}{\milli\meter}, 
OD=\SI{204}{\milli\meter}, 
complying with DIN 11866,
with a specified inner surface roughness 
of $R_\text{a} < \SI{0.8}{\micro\meter}$, 
made by Herrli AG in Kerzers, Switzerland, 
is sandwiched between the two UCN shutters.
The tube was electro-polished to obtain the final surface finishing.

Together with the shutters, 
this results in a cylindrical UCN storage volume 
of length 
\SI{1020.0\pm1.1}{\milli\meter},
with an inner diameter of \SI{200\pm0.5}{\milli\meter}, and therefore with
a volume of \SI{32044\pm164}{\centi\meter^3}.

The shutters are mounted in a support structure made from aluminum profiles, 
which provide a solid, flat pedestal and simplifies handling during assembly.

At the flanges of both shutters 
\SI{150\pm1}{\milli\meter}
long SST extension tubes are attached, 
which guarantee the free movement of the shutter blades and
provide connections to the beamport and the detector.
Two sizes of 
Cascade detectors \footnote{CD-T Technology, Hans-Bunte Strasse 8-10, 69123 Heidelberg, Germany}
were used for the measurements.
One large area 
('big')
Cascade UCN detector 
with 20\,cm $\times$ 20\,cm active area \cite{Goeltl2012}
and one 
with 10\,cm $\times$ 10\,cm active area 
('small detector') was employed.

\begin{figure}[ht]
\centerline{
\includegraphics[width=0.79\textwidth]{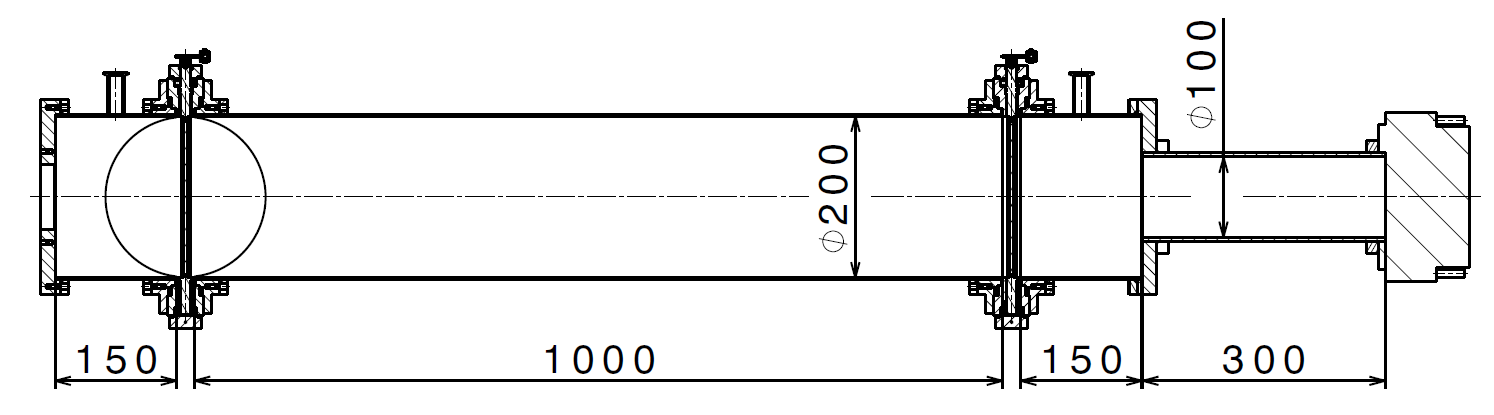}}
\caption{Technical design drawing of the storage experiment.
From left to right: connection flange to the beamport, 
connection guide with KF16 (2\,mm) flange, shutter~1, storage bottle, shutter~2, 
connection guide with KF16 flange,
connection flange, 
30\,cm horizontal extraction guide,
small Cascade detector.
The big Cascade counter was directly mounted 
onto the connection guide.
}
\label{BottleDrawing}
\end{figure}

\begin{figure}[ht]
 \begin{center}
\includegraphics[width=0.79\textwidth]{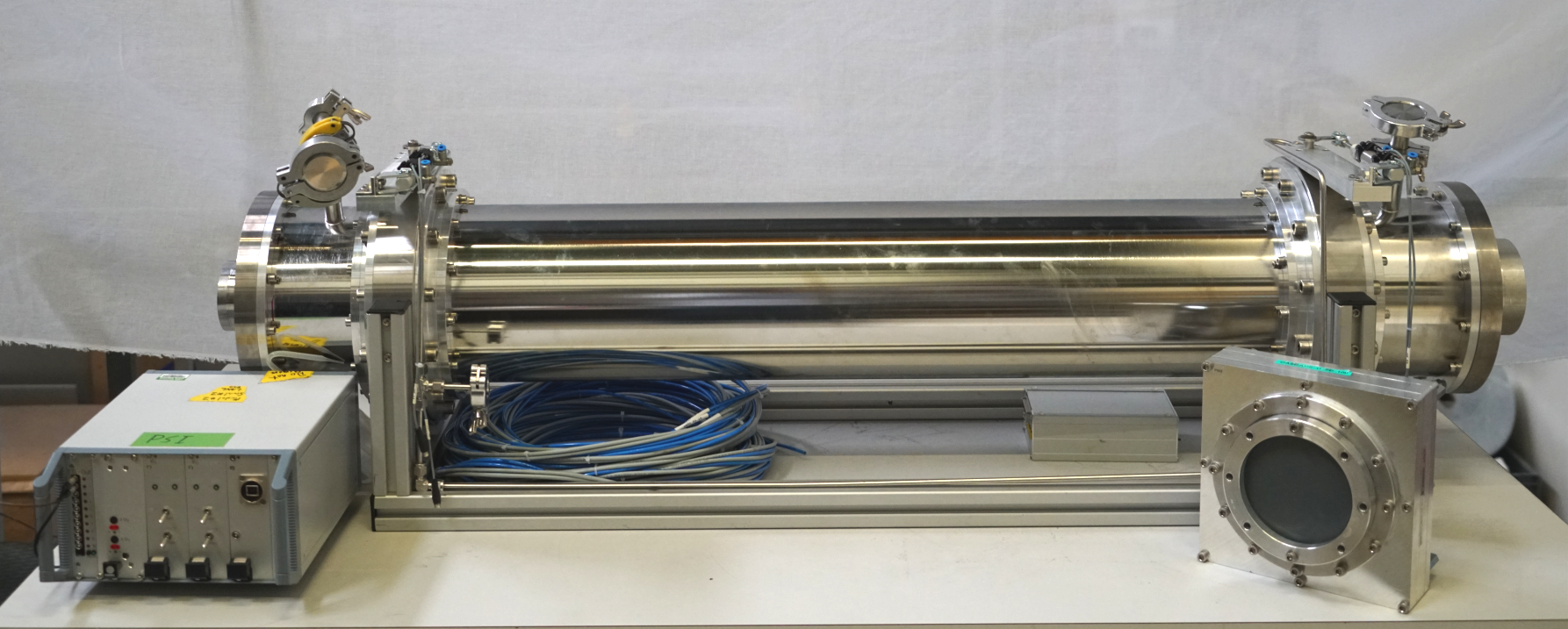}
%
\end{center}
\caption{%
The assembled UCN storage bottle
with the small Cascade detector on the right and
the timing control unit on the left.
}
\label{BottlePhoto}
\end{figure}

Two KF16 vacuum flanges
welded on the extension tubes 
provide ports for pumping and pressure measurement.
The one located upstream has the full opening of ID=\SI{16}{\milli\meter}, 
while the one before the detector has a reduced opening of \SI{2}{\milli\meter}.
Figures~\ref{BottleDrawing}~and~\ref{BottlePhoto} show a drawing
and a 
photo of the assembled storage bottle.

\subsection{Timing control}

The opening and closing sequence for the two shutters needs to 
be configurable and reproducible in order to perform UCN storage 
measurements.

5\,V TTL signals are used to 
control the air flow to the pneumatic actuators. 
The shutters 
go into the closed state when 
no signal (\SI{0}{\volt}) is present.

In the measurements described here, an Arduino Due board 
\footnote{https://www.arduino.cc/en/Main/ArduinoBoardDue}
is used 
for the timing logic. 
The board is based on a 
32-bit micro-controller running at a clock speed of \SI{84}{\mega\hertz}
and can be programmed using the C programming language 
and a hardware abstraction layer provided as open source software.
This makes it very
simple to implement a deterministic real-time capable digital controller. 
Due to the high clock speed, cycle times of \SI{100}{\micro\second} are 
easily accomplished.

The program flow is the following:

\begin{enumerate}

 \item A counter is being increased every \SI{100}{\micro\second} 
starting with the rising flank of the trigger input.

 \item In the first timing cycle after the trigger, the shutters are set to the filling
 position, shutter~1 (upstream) open, shutter~2 (downstream) closed.

 \item When the counter reaches the predefined filling time, both shutters are set
 to the closed position, and the storage time is started.

 \item When the counter reaches the sum of the predefined filling and storage times,
 the shutter~2 is opened and the counting period starts.

 \item When the counter reaches the sum of the predefined filling, storage and counting
 times, shutter~1 is opened again and shutter~2 closed;
 this is the only time when the storage vessel is being vacuum-pumped with open shutter.
 The counter is reset to zero and the trigger flag is set to false, waiting to be triggered again.

\end{enumerate}

All timing parameters are set through a serial connection from
a PC. 
Manual operation of the two shutters 
is also possible.

A custom-made \SI{10}{} channel galvanically-isolated TTL driver card is used
to convert the \SI{3.3}{\volt} output of the micro-controller board
to \SI{5}{\volt}.
The decoupled \SI{5}{\volt} outputs
then drive the TTL inputs of the two UCN shutters.


\section{Commissioning of the setup}


A set of measurements at the West-1 beamport at PSI was performed in
fall 2014.
The big
Cascade UCN detector 
was used in this measurement.
The commissioning measurements were done while the PSI source was 
operating with a proton beam pulse 
of only 3.2\,s length 
and a pulse length to repetition time ratio optimized
for the nEDM experiment
which is mounted at the South beamport.
These operating conditions were not well suited
for UCN density measurements at that time,
as they had to be done parasitically
to nEDM data taking.

\subsection{UCN transmission}

The UCN transmission of the storage bottle, 
the shutter, and
the connecting tubes was measured by comparing 2 setups.
After a first measurement with the  detector connected
directly to the beamport, the storage bottle was inserted between 
the detector and the beamport for a second measurement. 
The transmission is then calculated, 
similar as done in \cite{Blau2016},
as the ratio of counts 
obtained in the two measurements.
Measurements in both configurations were repeated several times. 
Since the UCN source output 
for different proton pulses
fluctuates more than standard counting statistics, 
the standard deviation of the mean was used as uncertainty. 

The proton pulse onto the UCN spallation target
in these measurements was always \SI{3.2}{\second} long 
with
a repetition time of \SI{340}{\second} 
at 2.2\,mA beam current.
Proton current variations 
are accounted for in the analysis.
%
%
UCN were detected continuously, but
a delayed time range for counting UCN was chosen
with a length of
230\,s
starting at
10\,s after the end of the proton pulse.
This guaranteed that only storable UCN were counted, as during the 
proton pulse also faster neutrons 
and other background events
may have been detected.

With the detector directly mounted to the 
beamport   
about \SI{1E7}{UCN} were counted per pulse
in the total time range 
and \SI{7.0E6}{UCN} in the delayed time window;
after mounting the storage bottle the respective UCN counts 
were about \SI{8E6}{UCN} (total window)
and \SI{5.7E6}{UCN} (delayed window).
Correcting for a small difference in proton beam current 
between the measurements,
this results in a UCN transmission
of \SI{81.7\pm0.1}{\percent}
caused by the Fermi potential and surface conditions of the storage bottle
and unavoidable small gaps 
between various UCN guiding parts
of the setup.

\subsection{UCN leakage}

\begin{figure}[h]
\begin{center}
 \includegraphics[width=0.79\textwidth]{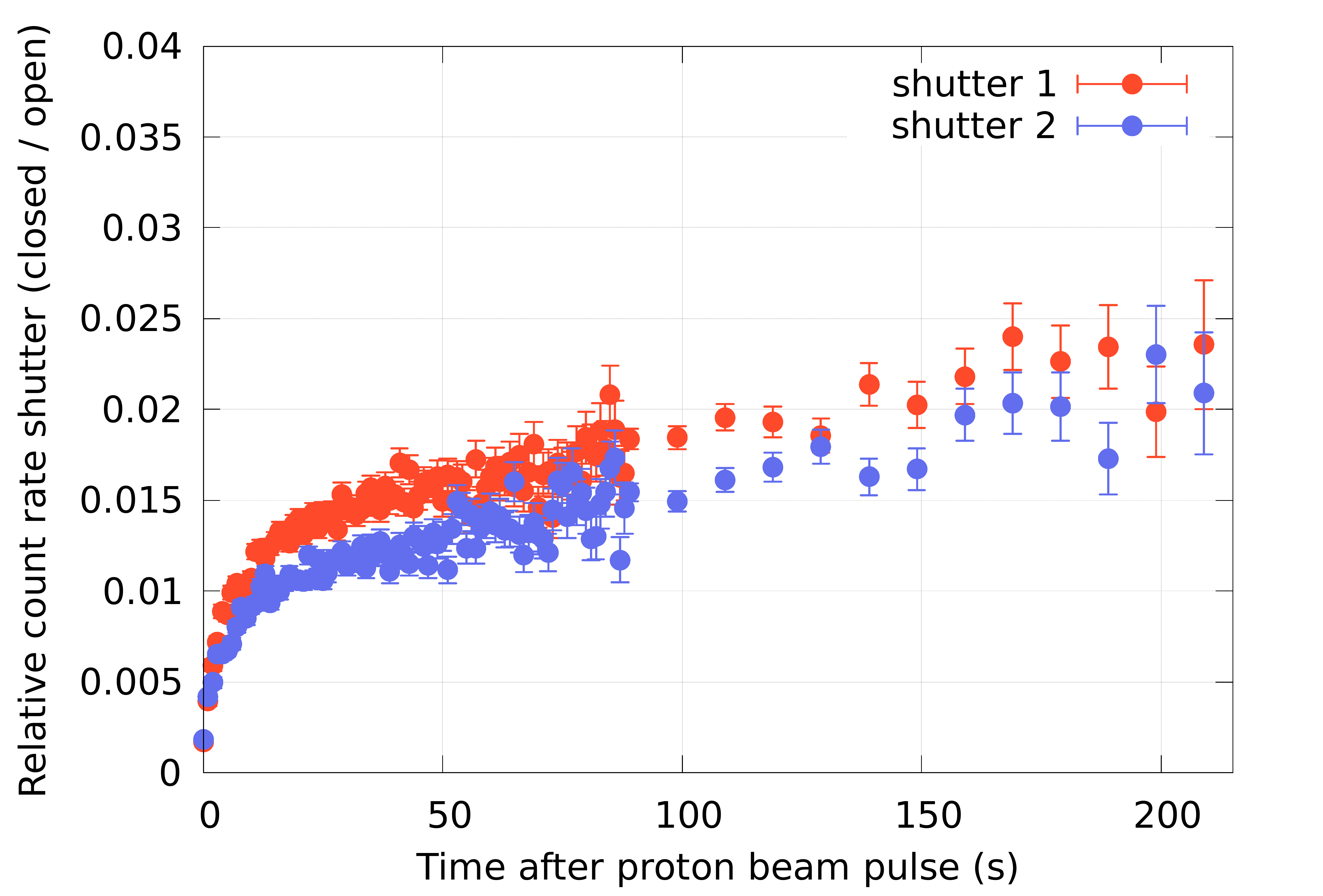}
 \caption{
UCN counts (leakage) versus time after the end of the proton pulse.
Bin-wise ratio between UCN counts with closed
and open shutters. 
Bins of size \SI{1}{\second} during the first \SI{90}{\second}, 
bins of size \SI{10}{\second} later.
The leakage rate rises quickly from \SI{0}{} to between \SI{1.0}{} and \SI{1.5}{\percent} after
the beginning of UCN production.
Afterwards, the leakage rate increases
to between \SI{2.0}{} and \SI{2.5}{\percent}, which is simply caused by the shorter emptying
time of the source with shutters opened.
 }
 \label{leakagerates}
\end{center}
\end{figure}

The two shutters are not completely tight with respect to UCN.
Comparing the UCN rate in the detector measured with both shutters open to 
the UCN rate with one shutter closed, one finds a UCN
leakage between \SI{1}{\percent} and \SI{2}{\percent} for both shutters.
Figure~\ref{leakagerates} shows the measured UCN leakage rates 
of the two shutters over time.

Fitting single-exponential decays to the 
count rates with different shutter positions
consistent with \cite{Goeltl2012},
results in emptying
times of the UCN source of 
\SI{36.0\pm.3}{\second} for both shutters open and
emptying times of \SI{40.0\pm.3}{\second} 
for shutter~1 closed 
and \SI{40.8\pm.3}{\second} for shutter~2 closed.
This change in emptying time leads to an apparent slow rise of the leakage rate 
over time.
The low leakage rate at short times after the end of the proton pulse 
is due to the 
obstruction of the direct UCN flight trajectories to the detector by the closed shutter, 
which causes a 
delayed UCN arrival.
%

\subsection{UCN storage}

\begin{figure}[h]
 \begin{center}
  \includegraphics[width=0.79\textwidth]{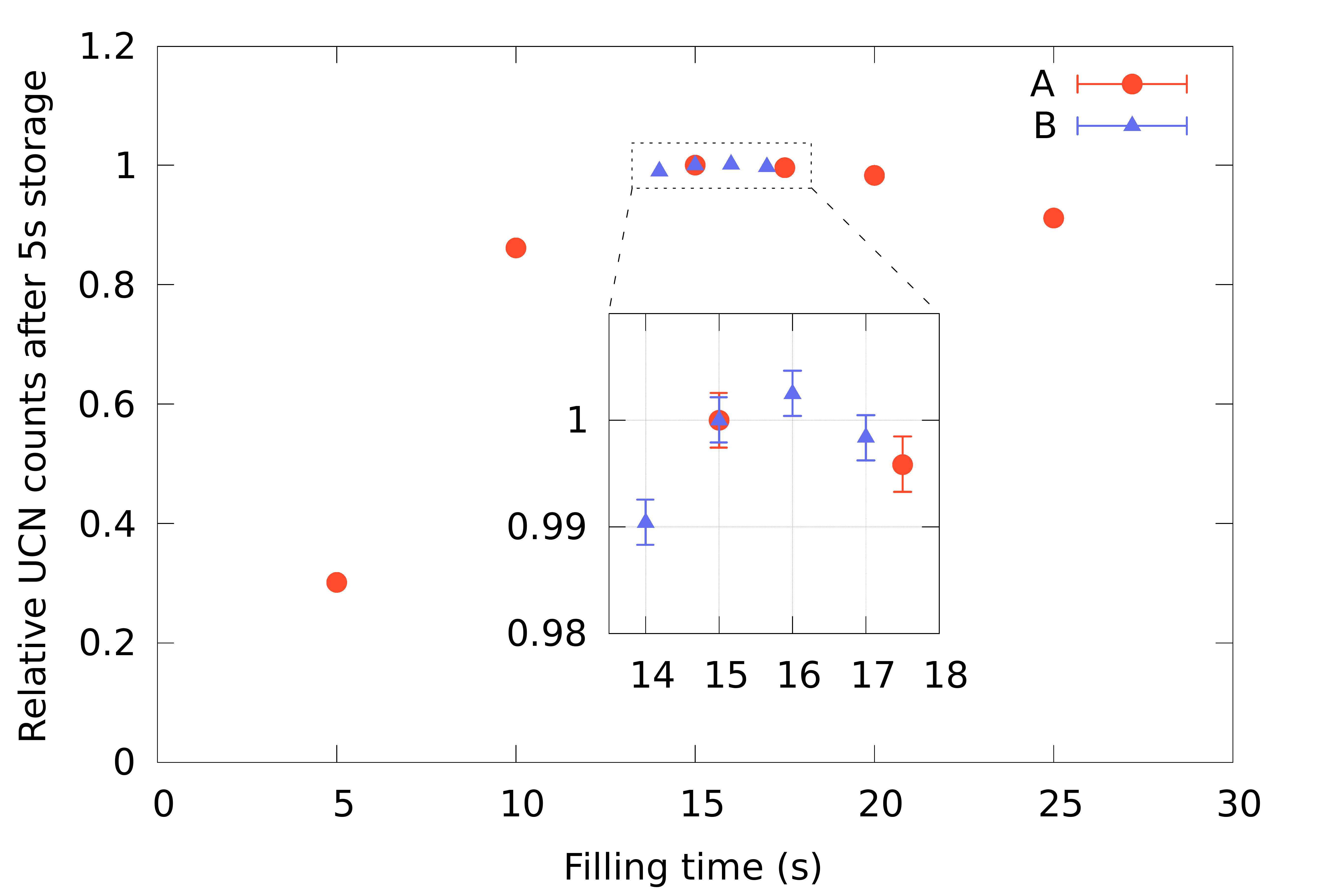}
  \caption{
	Relative UCN counts measured after \SI{5}{\second} of storage for different filling times.	
  The setup was dismounted between the two performed measurement periods A and B,
	with a different performance of the UCN source.
	The measurements are in good agreement. 
	Shown values are normalized to the value
  at $t=$\SI{15}{\second} measured in period A and B.
	The best output was measured for $t=$\SI{16}{\second}. 
	Error bars are statistical only and only 
  shown in the zoomed-in area.}
  \label{fillTimePlot}
 \end{center}
\end{figure}

The filling time of the bottle was optimized with 5\,s storage measurements.
Figure~\ref{fillTimePlot} shows the recorded filling time scan
with a relatively broad maximum at t=\SI{16}{\second}.
The timing was started by the proton beam trigger signal at t=\,0
which occurs 
\SI{1}{\second} before the proton beam hits the target.
Using the optimized filling time,
four measurements with storage times of 2, 5, 25, and 105 seconds
and a background measurements with shutter 1 closed 
were performed.

The UCN leakage of the shutters made a correction of the measured
counts necessary.
The UCN counts in the detector were integrated starting from the time when
the second shutter was opened until about \SI{260}{\second} after the 
end of the proton beam pulse. 
The leakage counts were determined in an measurement
which was identical except that 
shutter~1 stayed closed all the time.

\begin{figure}[h]
 \begin{center}
  \includegraphics[width=0.79\textwidth]{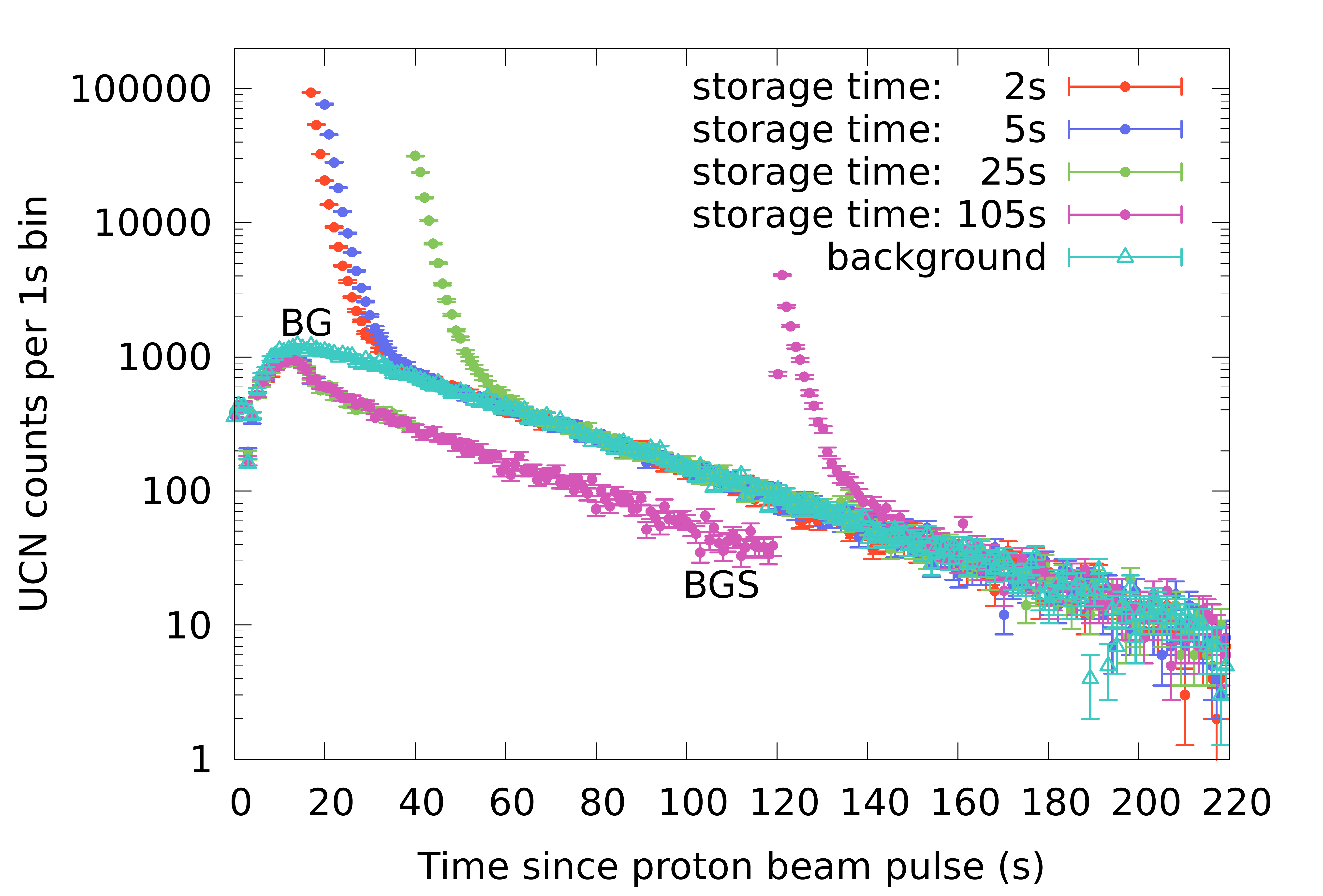}
  \caption{
	Time spectra of UCN counts in the storage measurement for different storage times using
  the big Cascade detector. 
	The background curve ('BG') with one shutter closed 
	(open triangles) 
	does not show an emptying peak.
The background curve with both shutters closed ('BGS') was measured in the 105\,s
storage measurement, where shutter 2 opens after this time and 
the stored UCN are detected in the counting peak.	
	}
  \label{timespectra}
 \end{center}
\end{figure}

Figure~\ref{timespectra} shows an overlay of the measured time
spectra in the detector. 
For late times the UCN counts 
reach again the level of the background.

\begin{figure}[h]
 \begin{center}
  \includegraphics[width=0.79\textwidth]{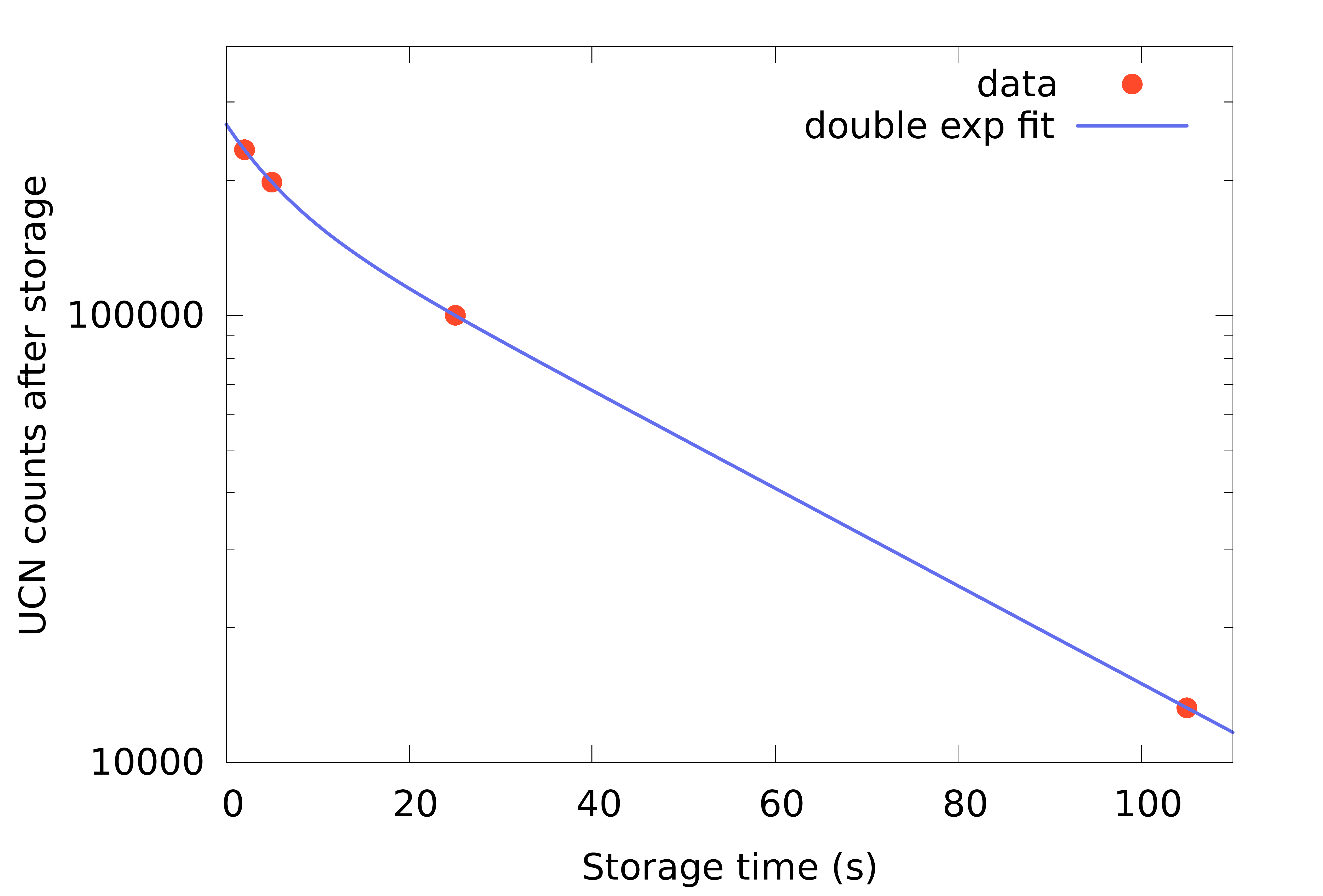}
  \caption{
	UCN counts in the big Cascade detector after different storage times (symbols) 
	together with 
	the fit function for a double-exponential fits
	drawn as a line (as given in Tab.\ref{fitparameter}).
}
  \label{NewBigStorageCurveLogscale}
 \end{center}
\end{figure}

Figure~\ref{NewBigStorageCurveLogscale} shows
the storage curve together with the fit from the leakage time spectrum
using a double-exponential fit function. 
Tab.\ref{fitparameter} lists the fit parameters.
The storage time constant
$\tau_2$
represents the main (slow) UCN component.
Statistical errors on the individual
points are well below \SI{1}{\percent} and are therefore smaller than the symbols
in the figure.

\begin{table}[htb]
\begin{center}
\begin{tabular}{c|c|c|c|c|c}
	{\small Detector}   &  $A_1$        &  $\tau_1$(s)         &  $A_2$        &  $\tau_2$(s)    & red.$\chi^2$     \\\hline
	Big        & \SI{30\pm7}{} &  \SI{6 \pm5}{}  & \SI{56\pm8}{}  &  \SI{40\pm8}{}& \SI{1.13}{}\\
  Small      & \SI{17\pm2}{} &  \SI{9 \pm2}{}  & \SI{52\pm2}{}  &  \SI{40\pm1}{}& \SI{1.23}{}\\
%
\end{tabular}
\caption{
Parameters from the double-exponential fit 
to the leaking UCN measured with the two Cascade detectors. 
}
\label{fitparameter}
\end{center}
\end{table}

Ultracold neutron densities can easily be calculated 
by dividing the UCN counts after a selected storage time 
with the total volume of the storage. 
Because of the described measurement conditions,
mainly due to the short proton pulse length, 
only half the UCN density compared
to previous measurements in Ref.~\cite{Goeltl2012}
was observed.
Dedicated UCN density measurements with the 
storage bottle described here, 
will be reported in a forthcoming publication.

\subsection{Small detector}
\label{small}

\begin{figure}[h]
 \begin{center}
  \includegraphics[width=0.79\textwidth]{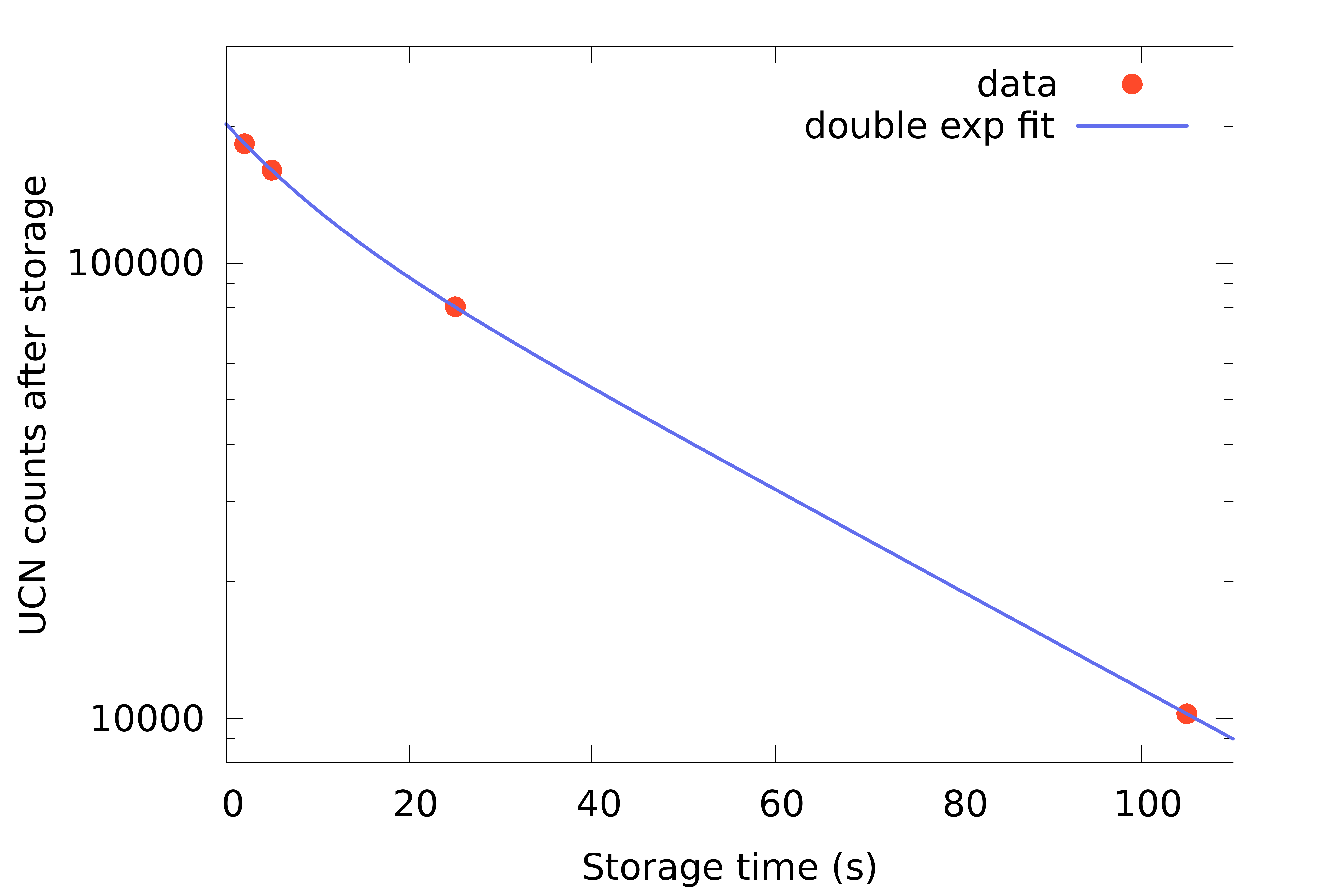}
  \caption{
		UCN counts in the small Cascade detector 
		after different storage times (symbols) 
	together with 
	the fit function for a double exponential fits drawn as 
	a line (as given in Tab.\ref{fitparameter}).
}
  \label{NewSmallStorageCurveLogscale}
 \end{center}
\end{figure}

The measurements were subsequently repeated 
and confirmed with the small
Cascade detector,
which was connected to the bottle using a straight 
\SI{300}{\milli\meter} piece of \isotope{NiMo}-coated 
acrylic tube
with \SI{100}{\milli\meter} inner diameter.
The measured storage curve is shown in Fig.~\ref{NewSmallStorageCurveLogscale}.
Again, using a double-exponential fit function
we obtain the results given in 
Tab.\ref{fitparameter}.
The smaller extraction opening and 
detector area show no significant influence on the 'relevant' slow component,
which represents the UCN storable for longer observation times.

\section{Summary}

We designed and constructed a storage experiment 
made from stainless steel components
in order to have a mobile and robust
setup with fast shutters,
to determine UCN densities at 
various beamports.
The storage bottle was successfully commissioned
at the West-1 beamport of the PSI UCN source. 
The UCN performance characteristics of the storage bottle,
namely
UCN transmission and storage times were determined
for the PSI UCN beam characteristics.
Measured integral UCN counts and the known bottle volume
can hence be used to derive UCN densities
for selected storage times.

\section{Acknowledgements}
This work is part of the Ph.D. thesis of Dieter Ries.
We would like to thank all people who contributed to design and construction 
of our setup:
M.~Horisberger,
M.~M\"ahr and the AMI shop, 
M.~M\"uller and his workshop,
as well as P.~R\"uttimann.
Support by the Swiss National Science Foundation
Projects 200020\_137664 and 200020\_149813 is gratefully acknowledged.
\hbadness=10000  


\end{document}